  \documentclass[amsmath,12pt,amssymb,preprint,prd,aps,nofootinbib]{revtex4}
  \usepackage{amsfonts} 
  \usepackage{graphicx} 
  \usepackage{epsfig}
  \usepackage{multirow}
  \usepackage{bm}
  \usepackage{color}
  \usepackage{amssymb}
  \usepackage{dsfont}

  \newcommand{\diracslash}[1]{#1\llap{/\kern2pt}}
  
  \newcommand{\be}{\begin{equation}}
  \newcommand{\ee}{\end{equation}}
  \newcommand{\bea}{\begin{eqnarray}}
  \newcommand{\eea}{\end{eqnarray}}
  \newcommand{\ba}[1]{\begin{array}{#1}}
  \newcommand{\ea}{\end{array}}
  
  \newcommand{\bt}{\begin{tabular}}
  \newcommand{\et}{\end{tabular}}

  \newcommand{\beas}{\begin{eqnarray*}}
  \newcommand{\eeas}{\end{eqnarray*}}

\begin{document}
\title{Thermodynamic properties and phase diagram of quark matter within non-extensive Polyakov chiral SU (3) quark mean field model} 
 \author{Dhananjay Singh}
\email{snaks16aug@gmail.com}
\affiliation{Department of Physics, Dr. B R Ambedkar National Institute of Technology Jalandhar, 
 Jalandhar -- 144008, Punjab, India}
 
\author{Arvind Kumar}
\email{kumara@nitj.ac.in}
\affiliation{Department of Physics, Dr. B R Ambedkar National Institute of Technology Jalandhar, 
 Jalandhar -- 144008, Punjab, India}

\def\be{\begin{equation}}
\def\ee{\end{equation}}
\def\bearr{\begin{eqnarray}}
\def\eearr{\end{eqnarray}}
\def\zbf#1{{\bf {#1}}}
\def\bfm#1{\mbox{\boldmath $#1$}}
\def\hf{\frac{1}{2}}
\def\kp{\zbf k+\frac{\zbf q}{2}}
\def\km{-\zbf k+\frac{\zbf q}{2}}
\def\hwo{\hat\omega_1}
\def\hwt{\hat\omega_2}

        \begin{abstract}
		
		In the present work, we apply Tsallis non-extensive statistics to study the thermodynamic properties and phase diagram of quark matter in the Polyakov chiral SU(3) quark mean field model. Within this model, the properties of the quark matter are modified through the scalar fields $\sigma, \zeta, \delta, \chi$, the vector fields $\omega, \rho$, $\phi$, and the Polyakov fields $\Phi$ and $\bar{\Phi}$ at finite temperature and chemical potential. Non-extensive effects have been introduced through a dimensionless parameter $q$ and the results are compared to the extensive case ($q\rightarrow1$). In the non-extensive case, the exponential in the Fermi-Dirac (FD) function is modified to a $q$-exponential form. The influence of $q$ parameter on the thermodynamic properties: pressure, energy, and entropy density as well as trace anomaly is investigated. The speed of sound and specific heat with non-extensive effects is also studied. 
 Furthermore, the effect of non-extensivity on the deconfinement phase transition as well as the chiral phase transition of $u, d,$ and $s$ quarks is explored. We found that the critical end point (CEP), which defines the point in the $(T - \mu)$  phase diagram where the order of the phase transition changes, shifts to a lower value of temperature, $T_{CEP}$, and a higher value of chemical potential, $\mu_{CEP}$, as the non-extensivity is increased, i.e., $q>$1.   
\end{abstract}
	
	\maketitle
	
	\section{\label{intro}Introduction}
	Quantum Chromodynamics (QCD) is a theory of strong interactions among the fundamental constituents of matter: quarks and gluons. In the low energy regime ($10^{-2}$ GeV), perturbative approaches to solve QCD become obsolete and we have to rely on non-perturbative methods. This led to the development of two theoretical approaches to QCD, namely lattice quantum chromodynamics (LQCD) and effective models of QCD. In LQCD, numerical techniques are applied on a discretized lattice of space and time. The discretization of space and time allows the equations to become computationally tractable to calculate the properties of hadrons and other QCD systems. Effective field models (EFMs) based on effective field theory were also developed to investigate the low-energy regime of QCD.  EFM uses a simpler set of degrees of freedom than those of the underlying theory. Studying thermodynamic properties and phase transition of quark matter at finite temperatures has always been a topic of great interest in high-energy physics. They play an important role in exploring physics shortly after the Big Bang \cite{big}. They are also relevant to studying the properties and structure of compact stars at high baryonic density and low temperature \cite{dex2014,baym,li2019,hinderer,weber}. At low baryonic density and high temperature, they are relevant to explain Quark Gluon Plasma (QGP) formation in relativistic heavy-ion collision \cite{marty}. Exploring the thermodynamic properties of dense quark matter is essential to understanding the nature of the phase transition in the ($T-\mu_{B}$) plane. The phase diagram of QCD contains information about the equilibrium phases of QCD and the physics of phase transition.
	
	\par Many theoretical models have been used to study the properties of hot and dense quark matter. These include Quark-Meson Coupling (QMC) model \cite{tsushima,saito}, Polyakov QMC (PQMC) model \cite{schaefer,herbst}, Nambu-Jona-Lasinio (NJL) model \cite{menezes,stiele}, Polyakov NJL (PNJL) model \cite{fukushima,ratii}, Linear Sigma Model (LSM) \cite{abu}, Polyakov LSM (PLSM) \cite{gatto, schaefer2009,kovacs,kovacs2022}, Chiral SU(3) Quark-Mean-Field (CQMF) model \cite{wang, wang2001}, and Polyakov CQMF (PCQMF) model \cite{manisha, nisha}. However, the statistical approach used in all these studies is often the standard Boltzmann-Gibbs (BG) statistics. The validation of BG statistics is limited to systems that satisfy the following conditions: short-range  type interactions, homogeneous and infinite heat bath, weak or no correlations between elements, and smooth boundary conditions. Violation of any or more of these conditions questions the validation of BG statistics.
	
	\par Thus, Tsallis proposed non-extensive statistics which depict a more realistic environment in heavy ion collision experiments. This formalism is an extension of the BG statistics \cite{tsallis}. In this framework the usual exponential function is replaced by a $q-$exponential function \cite{zhao2020,rozy2016},
	\begin{equation}
		\label{eq}
		exp_{q}(x) = [1+(1-q)x]^{\frac{1}{1-q}}
	\end{equation}
	where $q$ is a dimensionless parameter called the non-extensive parameter and it contains all possible effects violating the standard BG statistics. Correspondingly, the logarithmic function also gets modified to
	\begin{equation}
		\label{lnq}
		ln_{q}(x) = \frac{x^{1-q}-1}{1-q}.
	\end{equation}
	Both these functions return back to their usual forms when $q\rightarrow1$, i.e., $exp_{q}(x)\rightarrow exp(x)$, $ln_{q}(x)\rightarrow ln(x)$ and Tsallis statistics returns to BG statistics. As the $q$ parameter enters into the relevant formulae of the specific dynamical model it enables the phenomenological examination of the stability of the model under consideration against potential departures from the BG approach. It must be emphasized here that Tsallis statistics is not an alternative but a generalization of the BG statistics.
	
	\par The motivation for this research comes from recent non-extensive statistics studies which showed the violation of BG statistics in high energy physics. There is considerable agreement between experimental data and theoretical models based on non-extensive effects. As shown in \cite{bhat, asmaa}, BG statistics failed to reproduce the freeze-out parameters. Tsallis statistics has also been proven to accurately produce the transverse momentum distributions which follow a power law distribution rather than an exponential distribution as in the BG statistics \cite{wilk, li, marques, de,ryb}. Ref. \cite{combe} shows that Tsallis statistics can be used to explain the transport behaviour of cold atoms in optical lattices. Tsallis statistics also describes the behaviour of a system where ergodicity breaks down \cite{tirnakli, cirto}. Experimental data produced by PHENIX \cite{phenix}, STAR \cite{star} as well as RHIC collaborations by ATLAS \cite{atlas}, ALICE \cite{alice}, and collaborations at the LHC by the CMS \cite{cms} matches very well within the framework of Tsallis statistics. Several authors have summarized the validation of non-extensive effects in high-energy physics as well as in astrophysics \cite{bediaga,lavagno2000,lavagno2007,beck,wilk2000,lavagno2001,biro2005,biro2009,lavagno2002,alberico2008,quarati,carvalho,livadiotis,leubner}. More details and diverse applications of Tsallis statistics can be found in \cite{tsallis_2}.

	\par The purpose of this paper is to study a strongly interacting system whose dynamics are governed by non-extensive statistics. We investigate the difference between thermodynamic properties of quark matter when Tsallis statistics is applied and compare it with the BG statistics. In addition, we have also studied the phase transition of quark matter within the realm of non-extensive statistics. For this, we have generalized the PCQMF model to its non-extensive version. We explore situations where both chemical potential and temperature are non zero which stipulate the effect of Tsallis statistics on the whole ($T - \mu_{B}$) plane in the phase diagram. Other models, such as the QHD model, have also been generalized to their non-extensive version to study the properties of nuclear matter \cite{periera2007}. Non-extensive statistical effects on the properties of quark matter have also been studied within a generalized NJL model \cite{rozy2009}. Furthermore, a generalized version of the MIT bag model has also been used to study the QCD phase diagram and stellar matter properties \cite{cardoso}. In \cite{shen}, a non-extensive linear sigma model is used to study the chiral phase transition. Recently, the PNJL model has also been generalized to its non-extensive version to study the thermodynamic properties and transport coefficients \cite{zhao2020}, phase transition and critical exponents of QCD matter \cite{zhao2021} as well as baryon number fluctuations \cite{zhao2023}. Moreover, non-extensive statistics have also been used to study fluctuations and correlations in high-energy nuclear collisions \cite{alberico2000}. In Ref. \cite{drago}, non-extensive effects have been considered to study the relativistic nuclear equation of state. Ref. \cite{lavagno2010} included non-extensive effects to study the hadron to QGP transition. The bulk properties of protoneutron stars have also been investigated using the non-extensive formalism \cite{lavagno2011}. Non-extensive thermodynamics of nuclear matter, as well as a black hole, has been studied in \cite{megias}. Moreover, in Ref. \cite{tiwari} non-extensive effects have been used to study the transport properties of hadronic matter.
	
	\par This paper is organized as follows: In Sec. \ref{formalism}, we introduce the non-extensive version of the Polyakov chiral SU (3) quark mean field model ($q$-PCQMF). The impact of the non-extensive parameter, $q$, on various thermodynamic properties of quark matter is discussed in Sec. \ref{thermprop}. In Sec. \ref{phase}, we analyze in detail the influence of the parameter $q$ on the chiral and deconfinement phase transition at finite quark chemical potential and finite temperature. Finally, in Sec. \ref{summary}, we provide a brief summary of our work.

	\newpage
	\section{\label{formalism} Polyakov chiral SU(3) quark mean field model}
	Before presenting the $q$-PCQMF model, let us introduce the main concepts of the PCQMF model. This model uses quarks and mesons as degrees of freedom. At finite temperature and density, the interactions between quarks in this model are described by the exchange of scalar meson fields $\sigma$, $\zeta$, and $\delta$ and vector meson fields $\omega$, $\rho$ and $\phi$. The polyakov fields $\Phi$ and $\bar{\Phi}$ are introduced in the model to study the features of deconfinement phase transition. The attractive part of interaction is represented by the scalar meson fields while the repulsive interaction is through the vector meson fields. The non-strange scalar isoscalar meson field, $\sigma$, corresponds to the scalar meson of mass 417 MeV \cite{wang2001} and has ($u\bar{u}/d\bar{d}$) quark as its content. The scalar meson interacts with the strange quark through a strange scalar isoscalar field, $\zeta$ with ($s\bar{s}$) as its content and a mass of 1170 MeV \cite{wang2001}. The scalar isovector field, $\delta$ is introduced in the model to study the effect of isospin asymmetry and has ($u\bar{u}-d\bar{d}$) as its content. Furthermore, the model introduces the broken scale invariance through a scalar dilaton field, $\chi$ also known as glueball field \cite{schechter,gomm,heide,carter,ko}. 
	
	\par The effective Lagrangian of 2 + 1 flavour and 3 colour PCQMF model is as follows 
	\begin{equation}
		{\cal L}_{{\rm PCQMF}} \, = \, {\cal L}_{q0} \, + \, {\cal L}_{qm}
		\, + \,
		{\cal L}_M\,
		+ \, {\cal L}_{\Delta m} \, + \, {\cal L}_{h} \, -U(\Phi(\vec{x}),\bar{\Phi}(\vec{x}),T). \label{totallag}
	\end{equation}
	
	In the above equation ${\cal L}_{q0} = \bar \psi \, i\gamma^\mu \partial_\mu \psi$, represents the free part of massless quarks. The interactions between quark and meson is represented by
	\begin{eqnarray}
		{\cal L}_{qm}  =  g_s\left(\bar{\psi}_LM \psi_R+\bar{\psi}_RM^+\psi_L\right)- g_v\left(\bar{\psi}_L\gamma^\mu l_\mu \psi_L+\bar{\psi}_R\gamma^\mu r_\mu \psi_R\right), 
	\end{eqnarray}
	
	where $\psi=\left(u,d,s\right)$ and $g_v$$(g_s$) are vector(scalar) coupling constants. The spin-0 scalar ($\Sigma$) and pseudoscalar ($\Pi$) mesons are expressed as \cite{papa1999}
	\begin{equation}
		M(M^{\dagger}) = \Sigma \pm i\Pi = \frac{1}{\sqrt{2}}\sum_{a=0}^{8}(\sigma^a\pm i\pi^a)\lambda^a,
		\end{equation}
	where $\sigma^a (\pi^a)$ represent the scalar (pseudoscalar) meson nonets and $\lambda^a$ are the Gell-Mann matrices with $\lambda^0=\sqrt{\frac{2}{3}}\mathds{1}$. Similarly, spin-1 vector ($V_{\mu}$) and pseudovector ($A_{\mu}$) mesons are expressed as \cite{papa1999}
	\begin{equation}
		l_{\mu}(r_{\mu}) = \frac{1}{2}(V_{\mu}\pm A_{\mu}) = \frac{1}{2\sqrt{2}}\sum_{a=0}^{8}(v^a_{\mu}\pm a^a_{\mu})\lambda^a,
		\end{equation}
	with $v^a_{\mu}$ ($a^a_{\mu}$) being vector (pseudovector) mesons nonets.
	
	The meson interactions are described by the term ${\cal L}_M = {\cal L}_{\Sigma\Sigma} +{\cal L}_{VV} +{\cal L}_{SB}$. The self-interaction term for the scalar meson in the mean field approximation can be expressed as
 
	\begin{eqnarray}
            {\cal L}_{\Sigma\Sigma} =& -\frac{1}{2} \, k_0\chi^2
		\left(\sigma^2+\zeta^2+\delta^2\right)+k_1 \left(\sigma^2+\zeta^2+\delta^2\right)^2
		+k_2\left(\frac{\sigma^4}{2} +\frac{\delta^4}{2}+3\sigma^2\delta^2+\zeta^4\right)\nonumber \\ 
		&+k_3\chi\left(\sigma^2-\delta^2\right)\zeta 
		-k_4\chi^4-\frac14\chi^4 {\rm ln}\frac{\chi^4}{\chi_0^4} +
		\frac{d}
		3\chi^4 {\rm ln}\left(\left(\frac{\left(\sigma^2-\delta^2\right)\zeta}{\sigma_0^2\zeta_0}\right)\left(\frac{\chi^3}{\chi_0^3}\right)\right),
		\label{scalar0}
	\end{eqnarray}
 
	where $d = 6/33$, $\sigma_0$ and $\zeta_0$ corresponds to the vacuum value of $\sigma$ and $\zeta$ fields, respectively. These are expressed as $\sigma_0 = - f_\pi = 93$ MeV and $\zeta_0  = \frac{1}{\sqrt{2}} ( f_\pi - 2 f_K) = 115$ MeV, where $f_\pi$ is the pion decay constant and  $f_K$ is the kaon decay constant.
	
	The self-interaction term for the vector meson is written as
	\begin{equation}
		{\cal L}_{VV}=\frac{1}{2} \, \frac{\chi^2}{\chi_0^2} \left(
		m_\omega^2\omega^2+m_\rho^2\rho^2+m_\phi^2\phi^2\right)+g_4\left(\omega^4+6\omega^2\rho^2+\rho^4+2\phi^4\right). \label{vector}
	\end{equation}
	where the density-dependence of the mass of vector meson can be written as \cite{wang2003}
	\begin{eqnarray}
		m^2_\omega = m_\rho^2 =
		\frac{m_v^2}{1 - \frac{1}{2} \mu \sigma^2}\, ,
		\hspace*{.5cm} {\rm and} \hspace*{.5cm}
		m^2_\phi = \frac{m_v^2}{1 - \mu \zeta^2}\, .
	\end{eqnarray} 
	The parameters $m_v$ = 673.6 MeV and density parameter $\mu$ = 2.34 fm$^2$ are taken to yield ${m_\phi}$ = 1020 MeV and ${m_\omega}$ = 783 MeV.
	\par The masses of pseudo-scalar mesons arises due to the term \cite{wang2002,wang2004,papa1999}
	\begin{equation} 
		{\cal L}_{SB}=-\frac{\chi^2}{\chi_0^2}\left[m_\pi^2f_\pi\sigma + 
		\left(\sqrt{2}m_K^2f_K-\frac{m_\pi^2}{\sqrt{2}}f_\pi\right)\zeta\right].
		\label{esb_ldensity}
	\end{equation} 
	The exact mass of $s$ quark is generated through ${\cal L}_{\Delta m} = - \Delta m_s \bar \psi S \psi$, where $\Delta m_s = 29$ MeV and $S \, = \, \frac{1}{3} \, \left(I - \lambda_8\sqrt{3}\right) = {\rm diag}(0,0,1)$.
	
	\par In order to study the dependence of $T$ and $\mu_{B}$ in this model, we investigate the grand canonical potential density which can be expressed in the mean field approximation as 
	\begin{equation}
		\hspace*{-.4cm} 
		\Omega_{\rm{PCQMF}}= \mathcal{U}(\Phi,\bar{\Phi},T) - {\cal L}_M- {\cal V}_{vac} + \sum_{i=u,d,s}\frac{-\gamma_i k_BT}{(2\pi)^3}\int_0^\infty 
		d^3k\left\{ {\rm ln} 
		F^{-}+
		{\rm ln} F^{+}\right\},
        \label{tpdpcqmf}
	\end{equation}
	where, $\gamma_i$ = 2 is the spin degeneracy factor, and
	\begin{eqnarray}
		F^{-}=&1+e^{-3E^-}+3\Phi e^{-E^-}+3\bar{\Phi}e^{-2E^-}, \\
		F^{+}=&1+e^{-3E^+}+
		3\bar{\Phi} e^{-E^+}
		+3\Phi e^{-2E^+}.
	\end{eqnarray} 
	In the above equation, $E^- = (E_i^*(k)-{\nu_i}^{*})/k_BT$ and $E^+ = (E_i^*(k)+{\nu_i}^{*})/k_BT$ in which the effective single particle energy of quarks is given by $E_i^*(k)=\sqrt{m_i^{*2}+k^2}$. In Eq. (\ref{tpdpcqmf}) the term, ${\cal V}_{vac}$ is subtracted for the sake of getting zero vacuum energy. The in-medium quark chemical potential ${\nu_i}^{*}$ is linked to the chemical potential $\nu_i$ in free space by
	\begin{equation}
		{\nu_i}^{*}=\nu_i-g_\omega^i\omega-g_\phi^i\phi-g_\rho^i\rho,
		\label{mueff}
	\end{equation} 
	where, $g^i_{\omega}$, $g^i_{\phi}$ and $g^i_{\rho}$ are the coupling constants of vector meson fields with different quarks. The effective mass of quark, ${m_i}^{*}$, is given by   
	\begin{equation}
		{m_i}^{*} = -g_{\sigma}^i\sigma - g_{\zeta}^i\zeta - g_{\delta}^i\delta + {m_{i0}},
		\label{mbeff}
	\end{equation}
	where $g_{\sigma}^i$, $g_{\zeta}^i$ and $g_{\delta}^i$ describe the coupling strength of scalar fields with different quarks. The values of $g_{\sigma}^i$, $g_{\zeta}^i$ and  ${m_{i0}}$ are determined by fitting the vacuum masses of constituent quarks, $m_u,m_d,m_s$ = 313 MeV, 313 MeV, 490 MeV, respectively \cite{wang2003}.
	
	For the Polyakov loop effective potential, we have considered the commonly used logarithmic form given by \cite{fukushima,costa,roessner},
	\begin{eqnarray}
		\frac{U(\Phi,\bar{\Phi},T)}{T^4}&=&-\frac{a(T)}{2}\bar{\Phi}\Phi+b(T)\mathrm{ln}\big[1-6\bar{\Phi}\Phi+4(\bar{\Phi}^3+\Phi^3)-3(\bar{\Phi}\Phi)^2\big],
	\end{eqnarray}
	
	with the temperature dependent coefficients \cite{costa,roessner2008}:
	\begin{equation}\label{T}
		a(T)=a_0+a_1\bigg(\frac{T_0}{T}\bigg)+a_2\bigg(\frac{T_0}{T}\bigg)^2,\ \  b(T)=b_3\bigg(\frac{T_0}{T}\bigg)^3.
	\end{equation}
	The corresponding parameters $a_0$, $a_1$, $a_2$ and $b_3$ are given in Table \ref{tab:1}. 
	
	\begin{table}[h]
		\centering
		\begin{tabular}{|l|l|l|l|}
			\hline
			\multicolumn{1}{|c|}{$a_0$} & \multicolumn{1}{c|}{$a_1$} & \multicolumn{1}{c|}{$a_2$} & \multicolumn{1}{c|}{$b_3$} \\ \hline
			1.81                        & -2.47                      & 15.2                       & -1.75                      \\ \hline
		\end{tabular}
		\caption{Parameters in Polyakov effective potential}
		\label{tab:1}
	\end{table}
	
	In the pure gauge sector at vanishing chemical potential, the parameter $T_0=270$ MeV \cite{fukugita}. $T_0$ is rescaled from 270 MeV to 200 MeV when fermionic fields are included to compare the model results with existing lattice results \cite{manisha}. Finally, the solutions of the scalar fields $\sigma$, $\zeta$ and $\delta$, the dilaton field $\chi$, the vector fields $\omega$, $\rho$ and $\phi$ and the Polyakov field $\Phi$ and its conjugate $\bar{\Phi}$ are obtained by minimizing $\Omega$ with respect to these fields, $i.e.$,
	
	\begin{equation}
		\label{minimize}
		\frac{\partial\Omega}
		{\partial\sigma}=\frac{\partial\Omega}{\partial\zeta}=\frac{\partial\Omega}{\partial\delta}=\frac{\partial\Omega}{\partial\chi}=\frac{\partial\Omega}{\partial\omega}=\frac{\partial\Omega}{\partial\rho}=\frac{\partial\Omega}{\partial\phi}=\frac{\partial\Omega}{\partial\Phi}=\frac{\partial\Omega}{\partial\bar\Phi}=0.
	\end{equation}
 
        The model parameters which are summarized in Table \ref{tab:2}. can be estimated using the masses of $\pi, K$ mesons, average masses of $\eta$ and $\eta^{'}$ as well as the vacuum masses of $\sigma$, $\zeta$ and $\chi$ mesons.
	
	\begin{table}[b]
		\scriptsize{
		\centering
		\begin{tabular}{|c|c|c|c|c|c|c|c|c|c|}
			\hline
			$k_0$           & $k_1$          & $k_2$          & $k_3$         & $k_4$         & $g_s$         & $\rm{g_v}$          & $\rm{g_4}$           & $h_1$          & $h_2$                            \\ \hline
			4.94                 & 2.12                & -10.16              & -5.38              & -0.06              & 4.76               & 10.92               & 37.5                 & 0               & 0                                  \\ \hline
			$\sigma_0$ (MeV) & $\zeta_0$(MeV)  & $\chi_0$(MeV)   & $m_\pi$(MeV)  & $f_\pi$(MeV)  & $m_K$(MeV)    & $f_K$(MeV)     & $m_\omega$(MeV) & $m_\phi$(MeV)  & $m_\rho$( MeV)                   \\ \hline
			-93                  & -96.87              & 254.6               & 139                & 93                 & 496                & 115                 & 783                  & 1020                & 783                                   \\ \hline
			$g_{\sigma}^u$  & $g_{\sigma}^d$ & $g_{\sigma}^s$ & $g_{\zeta}^u$ & $g_{\zeta}^d$ & $g_{\zeta}^s$ & $g_{\delta}^u$ & $g_{\delta}^d$  & $g_{\delta}^s$ & $\rho_0$(fm$^{-3}$) \\ \hline
			3.36                 & 3.36                & 0                   & 0                  & 0                  & 4.76               & 3.36                & -3.36                & 0                   & 0.15                                  \\ \hline
			$g^u_{\omega}$  & $g^d_{\omega}$ & $g^s_{\omega}$ & $g^u_{\phi}$  & $g^d_{\phi}$  & $g^s_{\phi}$  & $g^u_{\rho}$   & $g^d_{\rho}$    & $g^s_{\rho}$   & $d$                              \\ \hline
			3.86                 & 3.86                & 0                   & 0                  & 0                  & 5.46               & 3.86                & -3.86                & 0                   & 0.18                                  \\ \hline
		\end{tabular}
		\caption{The list of parameters used in the present work.}
		\label{tab:2}
	}
	\end{table}
	
	\subsection{\textbf{\label{nepcqmf} q-PCQMF model}}
	\label{qpcqmf}
	The PCQMF model assumes the additivity of some thermodynamical quantities, like, entropy. For systems where long-range correlations and fluctuations are important, e.g., under phase transition, this is a very strong approximation. One way to take the nonadditivity of interacting systems into account is to consider a micro-canonical ensemble of Hamiltonian systems \cite{gross2004,gross2006}. An alternative approach is to make use of Tsallis statistics \cite{tsallis}. As discussed in Sec. (\ref{intro}), resigning from the assumption of additivity and describing theoretical models based on the non-extensive approach appears to be a promising possibility. In this spirit, we have extended the PCQMF model to its non-extensive version by using Tsallis statistics instead of BG statistics, which implies the replacements described in Eq.(\ref{eq}) and (\ref{lnq}). However, we have taken the following simplifications: Firstly, Polyakov loop potential experience non-extensive effects only implicitly and hence remain unchanged. Secondly, similar to the case of finite-size effects where the volume $V$ is treated on the same footing as $T$ and $\mu$ \cite{nisha}, the usual parameters of the PCQMF model are left undisturbed and the $q$ parameter is treated on the same footing as $T$ and $\mu$. 
 
	Hence, within the $q$-PCQMF model, the thermodynamic potential density becomes
	\begin{equation}
		\label{qtpd}
		\hspace*{-.4cm} 
		\Omega_{q}= \mathcal{U}(\Phi,\bar{\Phi},T) - {\cal L}_M- {\cal V}_{vac} + \sum_{i=u,d,s}\frac{-\gamma_i k_BT}{(2\pi)^3}\int_0^\infty 
		d^3k\left\{ {\rm ln_{q}} 
		F_{q}^{-}+
		{\rm ln_{q}} F^{+}_{q}\right\},
	\end{equation}
	where
	\begin{eqnarray}
		F_{q}^{-}=&1+e_{q}(-3E^-)+3\Phi e_{q}(-E^-)+3\bar{\Phi}e_{q}(-2E^-), \\
		F_{q}^{+}=&1+e_{q}(-3E^+)+
		3\bar{\Phi} e_{q}(-E^+)
		+3\Phi e_{q}(-2E^+).
	\end{eqnarray} 
	
	\par The coupled equations obtained by minimizing $\Omega_{q}$ in Eq. (\ref{qtpd}) are given in the appendix. The $q$ versions of vector density, $\rho_{q,i}$, and scalar density, $\rho_{q,i}^{s}$, of quarks are written as 
	\begin{equation}
		\rho_{q,i} = \gamma_{i}N_c\int\frac{d^{3}k}{(2\pi)^{3}}  
		\Big(f_{q,i}(k)-\bar{f}_{q,i}(k)
		\Big),
		\label{rhov0}
	\end{equation}
	and
	\begin{equation}
		\rho_{q,i}^{s} = \gamma_{i}N_c\int\frac{d^{3}k}{(2\pi)^{3}} 
		\frac{m_{i}^{*}}{E^{\ast}_i(k)} \Big(f_{q,i}(k)+\bar{f}_{q,i}(k)
		\Big),
		\label{rhos0}
	\end{equation}
	respectively, where $f_{q,i}(k)$ and $\bar{f}_{q,i}(k)$ are the $q-$modified Fermi-distribution functions for quark and anti-quark at finite temperature and are given by,
	\begin{equation}\label{qdistribution}
		f_{q,i}(k)=\frac{\Phi e^{q}_{q}(-E^-)+2\bar{\Phi} e^{q}_{q}(-2E^-)+e^{q}_{q}(-3E^-)}
		{[1+3\Phi e_{q}(-E^-)+3\bar{\Phi} e_{q}(-2E^-)+e_{q}(-3E^-)]^{q}} , 
	\end{equation}
	
	\begin{equation}\label{qdistribution1}
		\bar{f}_{q,i}(k)=\frac{\bar{\Phi} e^{q}_{q}(-E^+)+2\Phi e^{q}_{q}(-2E^+)+e^{q}_{q}(-3E^+)}
		{[1+3\bar{\Phi} e_{q}(-E^+)+3\Phi e_{q}(-2E^+)+e_{q}(-3E^+)]^{q}} .
	\end{equation}

	It is important to note that in the limit $q\rightarrow1,$ the standard Fermi-distribution functions are recovered and hence we return back to the standard (extensive) PCQMF model. 
	
	The following restriction must hold for $e_q(x)$ to always be a non-negative real function:
	\begin{equation}
		\label{condition}
		[1+(1-q)x]\ge 0
	\end{equation}
	In this paper, we consider only $q>1$. This is due to the fact that it is discovered that the typical value of the non-extensivity parameter $q$ for high-energy collisions is $1\le q\le 1.2$ \cite{marques,li2013,cleymans2013,azmi}. Furthermore, for $q>1$, $q-1$ describes intrinsic fluctuations of temperature in the system \cite{wilk2000,biro2005}, whereas, the interpretation of $q-1$ for $q<1$ is inconsistent \cite{kodama,garcia}.
	To comply with Eq.(\ref{condition}), for $q>1$, we can use the following Tsallis cutoff prescription \cite{tewel}
	\begin{equation}
		\label{cutoff}
		e_{q}(x)=
  \begin{cases} [1+(1-q)x]^{\frac{1}{1-q}}, \quad \text{for} \quad x\leq0,\\
			[1+(q-1)x]^{\frac{1}{q-1}}, \quad \text{for \quad} x>0 \quad.
   \end{cases}
	\end{equation}
	
	Also, let us observe that as $T\rightarrow 0$, as long as $q > 1$, $\Omega_{q} \rightarrow \Omega$ and the non-extensive effects vanish. This means that in studies where in the nuclear scale $T \approx 0$ (in the interior of neutron stars) we expect no non-extensive signature. Whereas, when $T$ is in MeVs, (heavy-ion collision experiments), one expects non-extensive effects to play an important role in the thermodynamic quantities. To study such effects, we apply the modified thermodynamic potential density Eq.(\ref{qtpd}) in Eq.(\ref{minimize}) to calculate the in-medium $q$-value of the vector and scalar fields. The isospin asymmetry can be incorporated through the definition $\eta=\frac{(\rho_d-\rho_u)}{(\rho_d+\rho_u)/3}$. The total baryon density is given by $\rho_B=\frac{1}{3} (\rho_u+\rho_d+\rho_s)$. The various chemical potentials are defined through
	$\mu_B=\frac{3}{2}(\mu_u+\mu_d)$, $\mu_I=\frac{1}{2}(\mu_u-\mu_d)$ and $\mu_S=\frac{1}{2}(\mu_u+\mu_d-2\mu_s)$, correspond to baryon, isospin and strangeness chemical potential, respectively.
	
	\section{\label{rd} Results and Discussions}
	\label{sec:4}
	In this section, we present various results on the thermodynamical properties of quark matter within $q-$extended Polyakov chiral $\text{SU(3)}$ quark mean-field model ($q$-PCQMF) presented in Sec. (\ref{qpcqmf}). In the $q$-PCQMF model, the effect of the $q$ parameter comes into the picture through the thermodynamic potential density, $\Omega_q$, which depends upon the $q-$ dependent scalar density, $\rho_{i}^{s}$ and vector density, $\rho_i$ of constituent quarks, which in turn depend upon the scalar fields  ($\sigma$, $\zeta$, $\delta$, $\chi$), the vector fields ($\omega$, $\rho$, $\phi$), and the Polyakov loop fields, ($\Phi$, $\bar{\Phi}$). As said before, these fields are calculated by solving the coupled system of non-linear equations. In Sect. \ref{thermprop}, the in-medium properties of the scalar and the Polyakov loop fields will be discussed. These fields will be used as input to understand the effect of $q$ on different thermodynamic properties of quark matter. We will discuss the behaviour of the susceptibility of these fields and study the $(T-\mu)$ phase diagram and the influence of the $q$ parameter on these quantities in Sect. \ref{phase}.
	
	\subsection{\label{thermprop} In-medium fields and thermodynamic properties}
	 
	\par
\begin{figure}
		\centering
		\includegraphics[width=15cm,height=12cm]{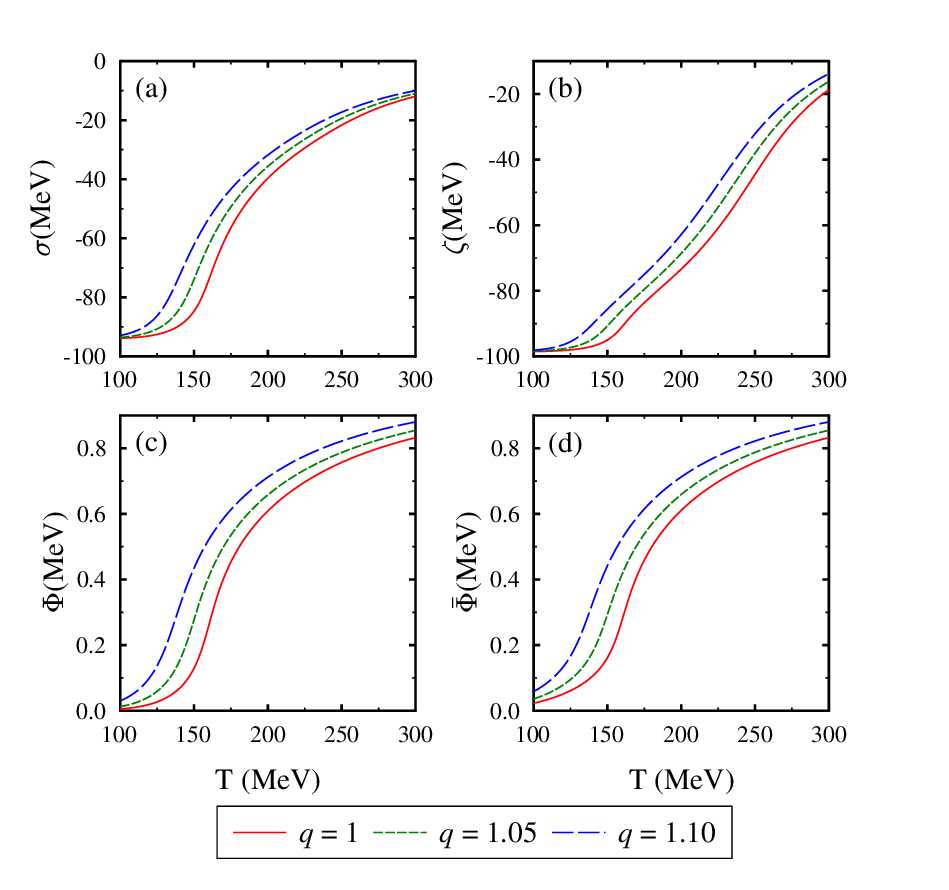}
		\caption{Scalar fields $\sigma$, $\zeta$ and Polyakov fields $\Phi$, $\bar{\Phi}$  plotted as a function of temperature T, for $q$ = 1, 1.05, and 1.10 at $\mu_B = 400$ MeV, $\mu_I = 40$ MeV, $\mu_S = 125$ MeV.}
		\label{fplots}
	\end{figure}

 In Fig. (\ref{fplots}), the variations of the scalar fields $\sigma$ and $\zeta$ as well as the Polyakov fields $\Phi$ and $\bar{\Phi}$ are shown as a function of temperature, $T$, for $q$ = 1, 1.05, and 1.10 at baryon chemical potential, $\mu_{B} = 400$ MeV, isospin chemical potential, $\mu_{I} = 40$ MeV and strangeness chemical potential, $\mu_{S} = 125$ MeV. We observe that the magnitude of $\sigma$ and $\zeta$ decrease with an increase in temperature. The temperature at which the magnitude of scalar fields begins to decrease rapidly is referred to as pseudo-critical temperature, $T_p$. This decrease in the magnitude of scalar fields may represent the restoration of chiral symmetry at high temperatures. Here, we can see that as $q$ increases, the value of $T_p$ decreases. As the temperature is increased the Fermi distribution function in Eqs. (\ref{qdistribution}) and (\ref{qdistribution1}) decreases resulting in a decrease in the magnitude of the scalar fields. At higher values of $q$, there is a considerable decrease in the magnitude of the fields even at lower temperatures. This may signify that the transition temperature is shifted towards a lower temperature value.
	\par In the mean-field approximation, the order parameters to study the deconfinement are the Polyakov loop fields, $\Phi$ and $\bar{\Phi}$. Variation in the magnitude of $\Phi$ and $\bar{\Phi}$ gives information about the deconfinement phase transition. As can be seen in Fig. \ref{fplots} (c) and (d), at lower temperatures, the value of both $\Phi$ and $\bar{\Phi}$ is approximately zero, indicating confined quarks within hadrons. With an increase in temperature, $\Phi$ and $\bar{\Phi}$ increase and reaches a critical temperature, where it jumps to a non-zero value, indicating the deconfinement phase transition. Similar to the scalar fields, we find that as non-extensivity increases (higher values of $q$), this increase in $\Phi$ and $\bar{\Phi}$ occurs earlier. This may indicate a decrease in the deconfinement temperature for $q>1$.
	
	\par One can calculate various thermodynamic properties such as energy density, pressure density, entropy density, and trace anomaly of quark matter with the help of the thermodynamics potential density, $\Omega_q$ in Eq. \ref{qtpd}. The pressure is given by
	\begin{equation}
		p_q(T)=-\Omega_q(T).
		\label{p1}
	\end{equation}
	The entropy density and energy density are given by
	\begin{equation}
		s_q(T)=-\frac{\partial \Omega_q(T)}{\partial T}, ~~~~~~ \epsilon_q(T)=\Omega_q(T)+\sum_{i=u, d, s} {\nu_i}^{*} \rho_i+Ts_q(T),
		\label{energy1}
	\end{equation}
	respectively.

\begin{figure}[h]
		\centering
		\includegraphics[scale=0.75]{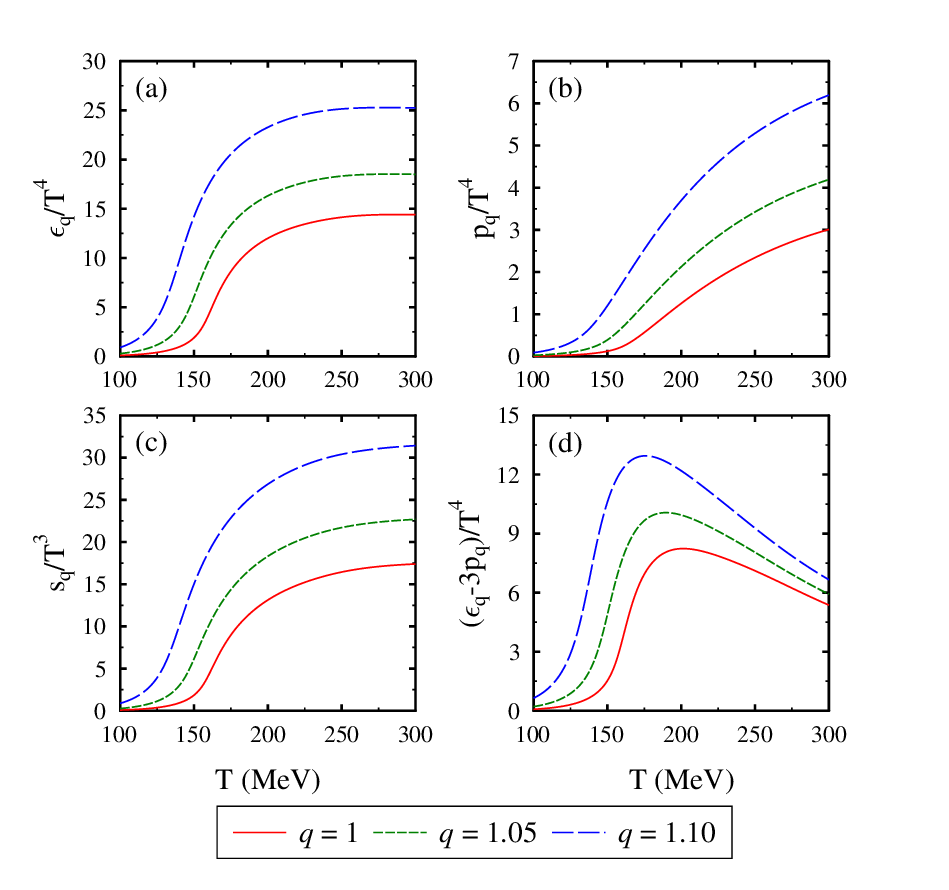}
		\caption{\centering Scaled energy density, $\epsilon_q/T^4$, pressure density, $p_q/T^4$, entropy density, $s_q/T^3$ and trace anomaly, $(\epsilon_q-3p_q)/T^4$ as a function of temperature $T$, for $q$ = 1, 1.05, and 1.10 at $\mu_B = 400$ MeV, $\mu_I = 40$ MeV, $\mu_S = 125$ MeV.}
		\label{therm}
	\end{figure}

	Fig. \ref{therm} shows the behaviour of $\epsilon_q/T^4$, $p_q/T^4$,  $s_q/T^3$ and 
	$(\epsilon_q-3p_q)/T^4$ at $\mu_{B} = 400$ MeV, $\mu_{I} = 40$ MeV, $\mu_{S} = 125$ MeV for $q = 1, 1.05$, and $1.10$. All these quantities show smooth behaviour implying a crossover phase transition. We see a sharp increase in the value of these quantities near the transition temperature. It is interesting to note that when $q = 1$, these thermodynamic quantities all tend to their Stefan Boltzmann (SB) limit at higher temperatures. However, with an increase in $q$, they increase rapidly and exceed their corresponding SB limit. For $\epsilon_q/T^4$ at $T = 300$ MeV, the value at $q = 1$ is 14.39 (very close to the SB limit, 15.627) but as $q$ is increased to 1.1, its value increases by $75\%$ to  25.24. For $s_q/T^3$ and $p_q/T^4$, these values are $80\%$ and $105\%$, respectively. This is the anticipated result in the $q-PCQMF$ model, because of the usage of Tsallis statistics instead of BG statistics. In Tsallis statistics, $q-1$ is interpreted as a deviation from the BG statistics \cite{biro20052, wilk2000}. The higher the value of $q-1$, the more the deviation from the BG statistics and correspondingly, the higher the value of the SB limit for quark matter. This is because when $T \rightarrow \infty $, the modified thermodynamic potential density doesn't return to its normal value, i.e., $\Omega_q \ne \Omega$. This means that for a system whose dynamics is described by Tsallis statistics, the high-temperature limit of various thermodynamic quantity is no longer the SB limit but a $q$-dependent Tsallis limit. This is similar to the results obtained in the $q$-NJL model \cite{zhao2020}. 
	
	\par
\begin{figure}[h]
		\centering
		\includegraphics[scale=0.75]{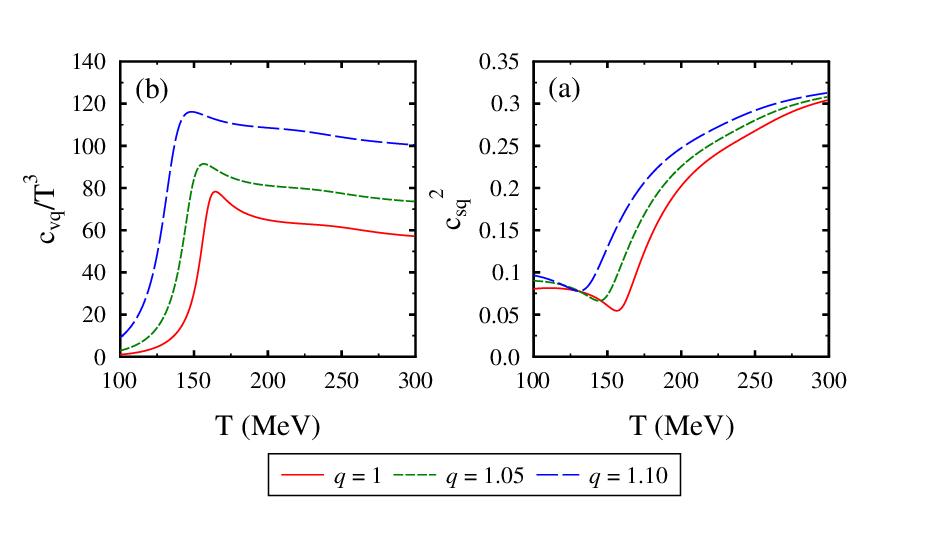}
		\caption{Scaled specific heat, $c_{vq}/T^3$ and the square of the speed of sound, $c_{sq}^2$ as a function of temperature $T$, for $q$ = 1, 1.05, and 1.10 at $\mu_B = 400$ MeV, $\mu_I = 40$ MeV, $\mu_S = 125$ MeV.}
		\label{soundplots}
	\end{figure}

 
	Some other relevant quantities in relativistic heavy-ion collision are the speed of sound squared at constant entropy $c_{sq}^2$ and specific heat at constant volume $c_{vq}$, which are defined as
	\begin{equation}
		c_{sq}^{2} = \left(\frac{\partial p_q}{\partial \epsilon_q}\right)_{s_{q}} = \frac{s_{q}}{c_{vq}}, \hspace{30pt}
		c_{vq} = \left(\frac{\partial \epsilon_q}{T}\right)_V.
	\end{equation}
	From Fig. \ref{soundplots} (a), we can see that the value of $c_v/T^3$ shows a similar trend to the other thermodynamic quantities discussed above. At $q = 1$, its value increases with an increase in temperature reaches a maximum near the pseudo-critical temperature and then tends to the usual SB limit, 63.14. Besides, for higher values of $q$, it also reaches a new $q$-dependent Tsallis limit.  In Fig. \ref{soundplots} (b), at $q = 1$, the value of $c_s^2$ dips near the pseudo-critical temperature but then approaches the SB value, 0.33. However, as $q$ increases, the dip disappears and the criticality vanishes. It is interesting to note that at high temperatures, both $s_q/T^3$ and $c_{vq}/T^3$ show similar growth. Therefore, non-extensive effects such as high-temperature limit are not observed in $c_{sq}^2 = \frac{s_q}{T^3}/\frac{c_{vq}}{T^3}$. Similar results are obtained in Ref. \cite{khuntia}. 
	
	\subsection{\label{phase} $T - \mu$ phase diagram}
	\par In this subsection, we study the $T-\mu$ phase diagram within the $q$-PCQMF model. We analyse three phase transitions in the present work: the chiral phase transition from the chirally broken phase to the chiral restoration phase for up and down quarks, the chiral phase transition for the strange quark, and the deconfinement phase transition. The effect of the $q$ parameter on the nature of these phase transitions is studied with the help of the susceptibilities, defined as
	\begin{equation}
		\chi_{\sigma_x} = \left(\frac{\partial \sigma_x}{\partial T}\right)_\mu, \hspace{25pt}
		\chi_{\sigma_y} = \left(\frac{\partial \sigma_y}{\partial T}\right)_\mu, \hspace{25pt}
		\chi_{\Phi} = \left(\frac{\partial \Phi}{\partial T}\right)_\mu,
		\hspace{25pt}
		\chi_{\bar{\Phi}} = \left(\frac{\partial \bar{\Phi}}{\partial T}\right)_\mu,
		\label{chis}
	\end{equation}
    where $\sigma_x$, for $u$ and $d$ quarks, and $\sigma_y$, for $s$ quark, are the scalar quark condensates which can be expressed in terms of scalar fields $\sigma, \zeta, \delta,$ and $\chi$ as \cite{manisha},
        \begin{equation}
            \sigma_x = \sigma_{u,d} = \frac{1}{m_{u,d}}\left(\frac{\chi}{\chi_{0}}\right)^2\Bigg[\frac{1}{2}m^2_{\pi}f_{\pi}(\sigma\pm\delta)\Bigg],
            \label{sigmax}
        \end{equation}
        \begin{equation}
            \sigma_y = \frac{1}{m_{s}}\left(\frac{\chi}{\chi_{0}}\right)^2\Bigg[\left(\sqrt{2}m^2_{k}f_{k} - \frac{1}{\sqrt{2}}m^2_{\pi}f_{\pi}\right)\zeta\Bigg].
            \label{sigmay}
        \end{equation}
	
	\begin{figure}
		\includegraphics[scale=0.80]{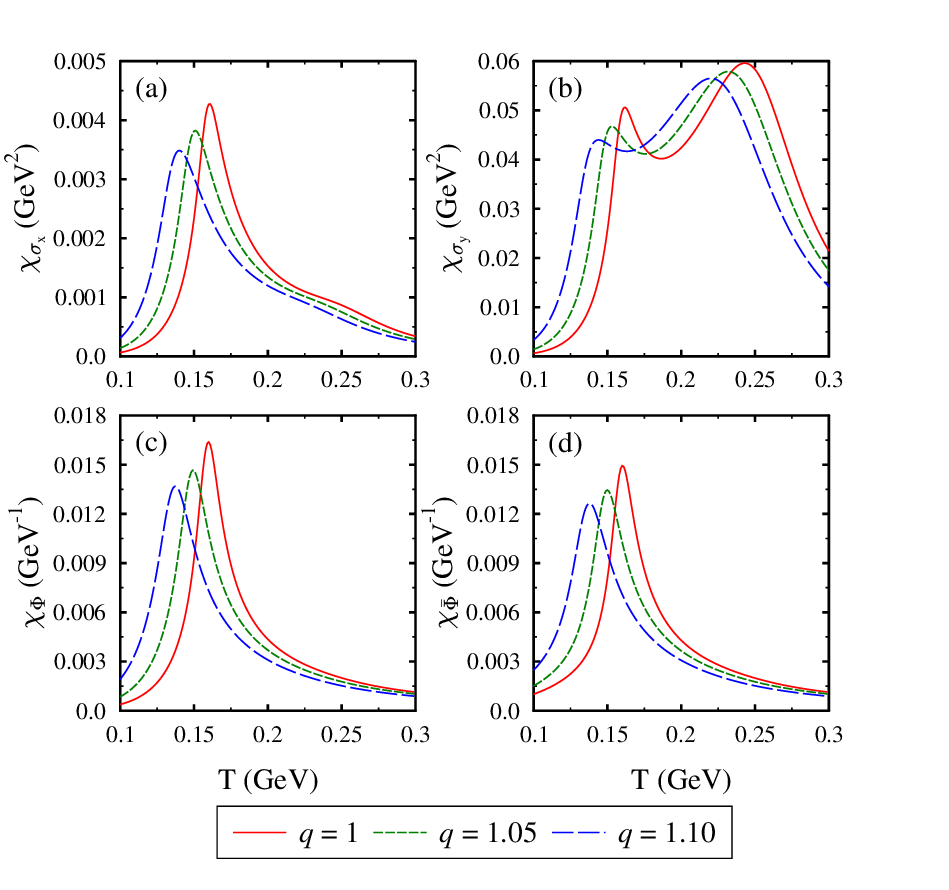}
		\caption{Susceptibilities of $\sigma_x$, $\sigma_y$, $\Phi$, and $\bar{\Phi}$ at $q$ = 1, 1.05, and 1.10 for $\mu_{B} = 400$ MeV, $\mu_{I} = 40$ MeV, $\mu_{S} = 125$ MeV.}
		\label{chiplots}
	\end{figure}
	
	From Fig. (\ref{fplots}) it is evident that there is a smooth transition of the order parameters (scalar fields ($\sigma$ and $\zeta$) and Polyakov loop ($\Phi$ and $\bar{\Phi}$)). We do however see a change in the behaviour of these order parameters as we increase the temperature. This is evidence of a crossover phase transition and we can study the susceptibilities defined in Eq. (\ref{chis}) to figure out the position of the pseudo-critical point $T_p$. The peak value of the susceptibilities $\chi_{\sigma_x}, \chi_{\sigma_y}$ corresponds to the pseudo-critical temperatures $T^q_{\chi}$, $T_\chi^{s}$ for chiral phase transition of up/down and strange quarks, respectively. The pseudo-critical temperature for the deconfinement phase transition $T_d$ is obtained through $\chi_{\Phi}$. Fig. (\ref{chiplots}) shows the variation of susceptibilities $\chi_{\sigma_x}$, for up and down quarks, $\chi_{\sigma_y}$, for the strange quark, and the Polyakov loop, $\chi_{\Phi}$ and $\chi_{\bar{\Phi}}$ at $\mu_{B} = 400, \mu_{I} = 40, \mu_{S} = 125$ MeV for $q = 1, 1.05$, and $1.10$. For $q$ = 1, $\chi_{\sigma_x}$ show a peak at $T \approx 160$ MeV, whereas, $\chi_{\sigma_y}$ has two peaks, one at $T \approx 160$ MeV and another around $T \approx 247$ MeV. Furthermore, only a single peak which coincides with the peak of $\chi_{\sigma_x}$ at $T \approx 160$ MeV is observed for $\chi_{\Phi}$ and $\chi_{\bar{\Phi}}$ at $q = 1$. The peak of the susceptibility gives us the pseudo-critical temperature. In case of two peaks in susceptibilities, the criteria for deciding the pseudo-critical temperature for chiral phase transition is given by $\sigma_{x,y}(T)/\sigma_{x,y}(T=0) < 1/2$, and for the deconfinement phase transition is $\Phi(T)/\Phi(T\rightarrow \infty) > 1/2$ \cite{mao}. It is evident from Fig. (\ref{chiplots}) that the value of these pseudo-critical temperatures decreases with increasing $q$ for both the chiral phase transition $T_\chi^q$ and the deconfinement phase transition $T_d$. It is also interesting to note that the chiral phase transition for the strange quark occurs at a much higher temperature as evidenced by the second peak at $T_\chi^s \approx 247$ MeV, whereas for lighter up and down quark it occurs at $T_\chi^q \approx 160$ MeV.
	
	\begin{figure}
		\includegraphics[scale=0.85]{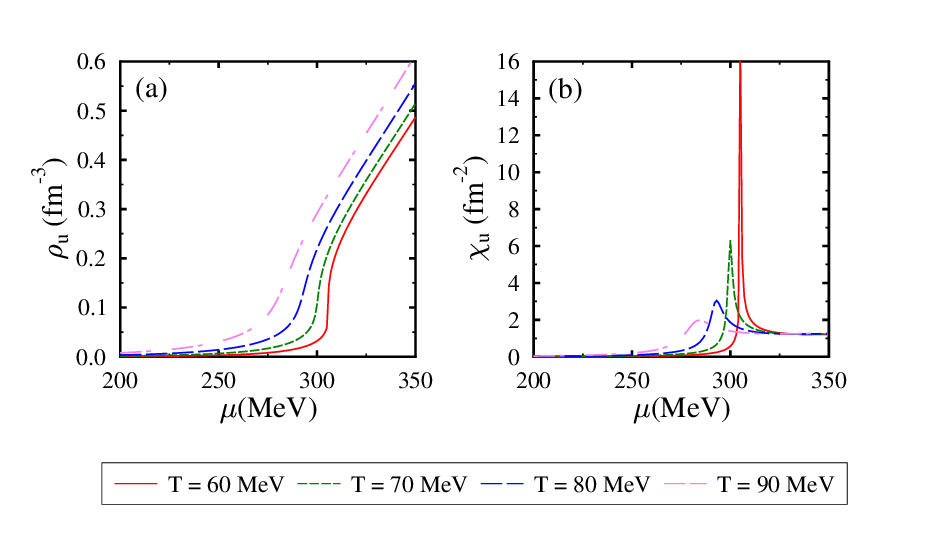}
		\caption{Quark number density $\rho_u$ and susceptibility $\chi_u$ of up quark as a function of quark chemical potential at T = 70, 80, 90, and 100 MeV for $q = 1$.}
		\label{rho_u1}
	\end{figure}
	
	\begin{figure}[h]
		\includegraphics[scale=0.85]{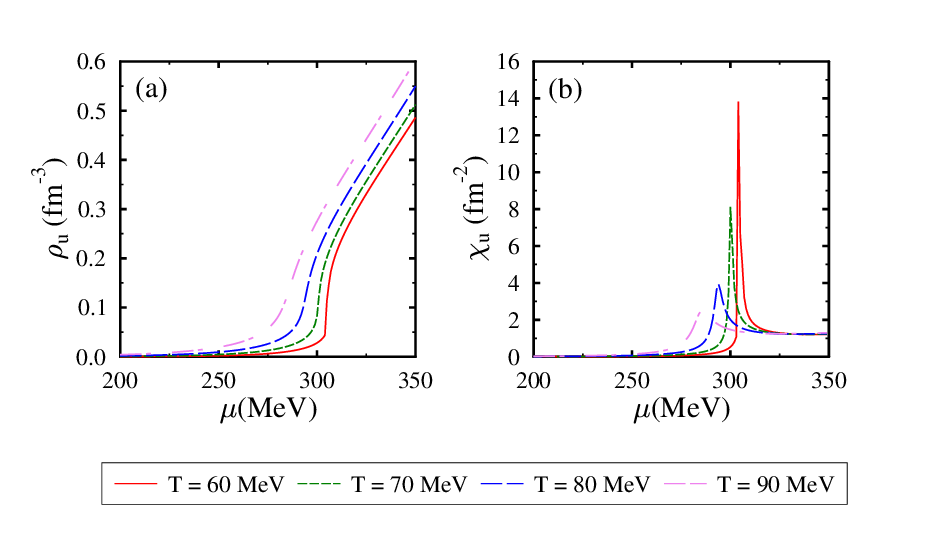}
		\caption{Quark number density $\rho_u$ and susceptibility $\chi_u$ of up quark as a function of quark chemical potential at T = 70, 80, 90, and 100 MeV for $q = 1.05$.}
		\label{rho_u105}
	\end{figure}

\begin{figure}[h]
	\includegraphics[scale=0.85]{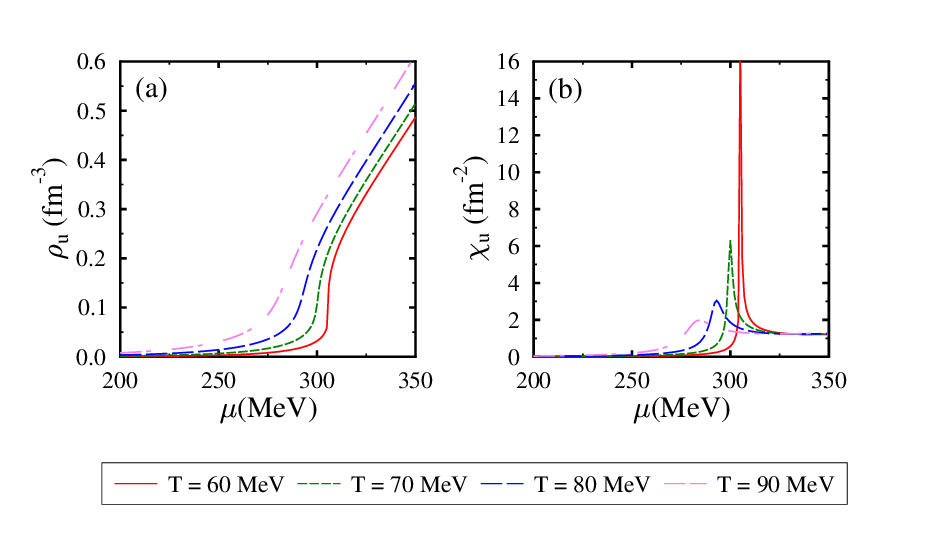}
	\caption{Quark number density $\rho_u$ and susceptibility  $\chi_u$ of up quark as a function of quark chemical potential at T = 70, 80, 90, and 100 MeV for $q = 1.10$.}
	\label{rho_u110}
\end{figure}
	
	\par The nature of the phase transition at a higher value of quark chemical potential is explored using quark number susceptibility. The chiral phase transition changes from a crossover to a first-order phase transition at higher values of $\mu_{q}$ and this can be seen by studying the discontinuity in the quark number density, $\rho_u$, or divergence in the corresponding susceptibility, $\chi_u$ defined as  
	\begin{equation}
		\chi_{u}=\left(\frac{\partial \rho_{u}}{\partial \mu}\right)_T.
	\end{equation}
	
	\par 
	In Fig. \ref{rho_u1} (a), quark number density is plotted as a function of quark chemical potential for different values of temperature and zero vector interaction, at $q$ = 1. It can be seen that when $T \le T_{CEP}$, $\rho_u$ shows a discontinuity, which can be seen in the corresponding susceptibility, $\chi_u$, plotted in Fig. \ref{rho_u1} (b), indicating the presence of first-order phase transition. When the temperature rises above $T_{CEP}$, the discontinuity in $\rho_u$ disappears suggesting the order of the phase transition to be a crossover. This change in the nature of $\rho_u$ can be used to determine the location of the CEP. For $q$ = 1, we observed the position of the CEP at around $(T,\mu)$=(90 MeV, 286 MeV) \cite{manisha}. When the value of $q$ is increased to 1.05, the peak in the $\chi_u$ is shifted towards higher $\mu$ and lower $T$ as shown in Fig. \ref{rho_u105}. For $q$ = 1.05, the position of the CEP is around $(T,\mu)$=(84 MeV, 290 MeV). A similar trend is shown in Fig. (\ref{rho_u110}) for $q = 1.10$, where we found the CEP at around $(T,\mu)$=(78 MeV, 293 MeV).

\begin{figure}
		\centering
		\includegraphics[width=10cm,height=8cm]{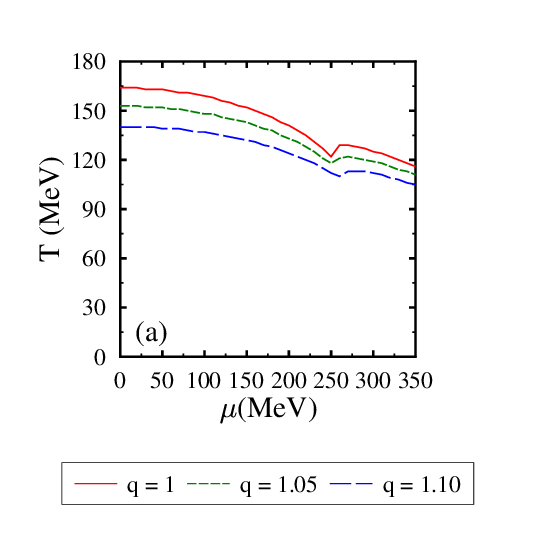}
		\caption{Deconfinement phase transition of quark matter at $\mu_S$ = $\mu_I$ = 0 MeV for $q$ = 1, 1,05, and 1.10.}
		\label{deconf}
	\end{figure}

\begin{figure}
		\centering
		\includegraphics[width=15cm,height=8cm]{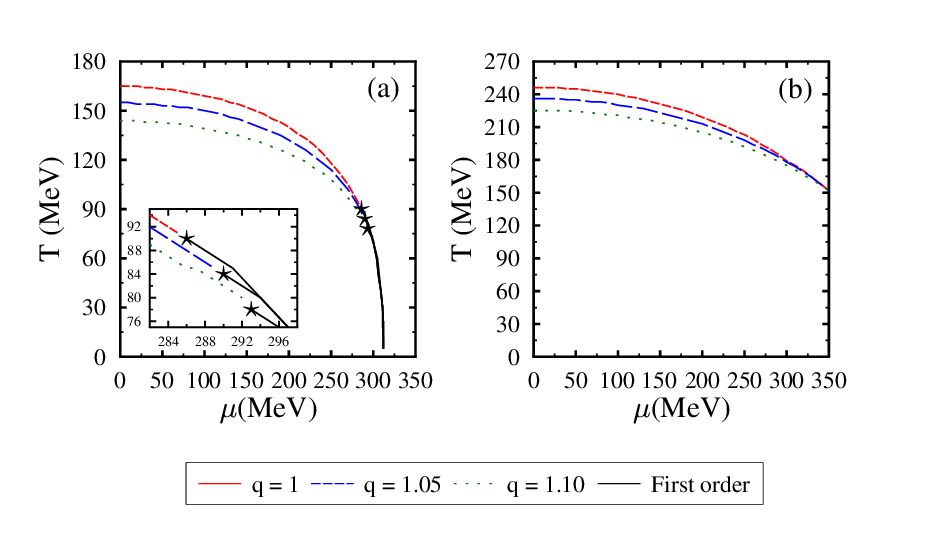}
		\caption{Chiral phase transition: (a) for $u$ and $d$ quarks, (b) for $s$ quark, at $\mu_S$ = $\mu_I$ = 0 MeV for $q$ = 1, 1,05, and 1.10.}
		\label{chiral}
	\end{figure}

	Fig. (\ref{deconf}) shows the deconfinement phase transition at $\mu_S = \mu_I = 0$ for different $q$ values. This deconfinement phase transition remains a crossover over the entire range of the quark chemical potential. As the value of $q$ increases, the phase transition temperature decreases to lower values. This means that with increasing non-extensivity, quarks can become deconfined even at lower temperatures. For $q$ = 1, at zero chemical potential, the value of this temperature, $T_d$ is 164 MeV. This value decreases to $T_d$ = 153 MeV for $q$ = 1.05 and $T_d$ =  140 MeV for $q$ = 1.10. However, the decrease is less significant at a higher value of $\mu$. For example, at $\mu=350$ MeV, the values of $T_d$ are observed to be 116, 111, and 105 MeV. This signifies that the impact of non-extensivity vanishes as we move towards lower temperatures.	In Fig. \ref{chiral} (a), the chiral phase transition of up and down quark is shown for the $q$-PCQMF model in the $T - \mu$ plane for different values of $q$, at $\mu_S = \mu_I = 0$. For plotting the phase boundary in the crossover region, we have used the maximum of $\chi_{\sigma}$ for a given value of $\mu$, whereas, for the first-order transition we have used the maximum of $\chi_u$ for a given value of $T$. With an increase in non-extensivity of the system, the regime of first-order phase transition becomes smaller and the CEP is achieved at a higher value of $\mu$ and lower value of $T$. It is also clear from Fig. \ref{chiral} (a) that the crossover transition, which occurs at higher temperatures is distinctly different but the first-order phase transition, which occurs at lower temperatures is similar when $q>1$. This is again evidence of the fact that non-extensivity vanishes at lower temperatures, i.e., $\Omega_{q}\rightarrow\Omega$ when $T\rightarrow0$. The chiral phase transition of the s quark is shown in Fig. \ref{chiral} (b), for $q$ = 1, 1.05, and 1.10. Contrary to the chiral phase transition of the u and d quark, this phase transition remains a crossover over the entire range of $(T - \mu)$ phase diagram. Besides, this phase transition also occurs at lower temperature values as $q$ becomes greater than one. However, this difference in temperature at different $q$ values disappears towards lower temperatures or higher chemical potentials indicating the vanishing of non-extensive effects.

	\newpage
	\section{\label{summary} Summary and Conclusions}
	
	To summarize, we have extended the PCQMF model to its non-extensive version and studied the properties of quark matter at finite temperature and density. We have studied the thermodynamic properties and phase transitions with Tsallis statistics and compared it to the usual BG statistics. The non-extensivity is introduced into our model through the thermodynamic potential density which modifies the form of the scalar and the vector densities of the quark resulting in the modification of the scalar, vector as well as polyakov loop fields. These modified fields are then further used to calculate various thermodynamic quantities. We have found that the quantities such as $\frac{\epsilon_q}{T^4}, \frac{p_q}{T^4}, \frac{s_q}{T^3},$ and $\frac{(\epsilon_q-3p_q)}{T^4}$ all tend towards a new $q$-related Tsallis limit rather than their usual SB limit at high temperature for $q>1$. We found that as $q$ increases, the criticality of $c_v/T^3$ and $c_s^2$ gradually decreases. Also, the high-temperature limit of $c_s^2$ is unaffected by the $q$ parameter due to a surprising cancellation \cite{zhao2020}. Regarding the influence of the $q$ parameter on the deconfinement phase transition, we found that this phase transition occurs at a lower value of $T$ as the $q$ parameter is increased. However, the nature of this phase transition is still a crossover, independent of $q$. Further, we have also studied the influence of the $q$ parameter on the chiral phase transition for the $u$ and $d$ quarks as well as for the s quark. We found that for $u$ and $d$ quarks, the nature of this phase transition changes order from a crossover at high temperatures to a first-order phase transition at lower temperatures. Also, as $q$ increases the CEP is shifted towards a lower $T$ and a higher $\mu$ value. However, for the s quark, the phase transition remains a crossover but is shifted towards lower $T$ values as $q$ becomes greater than one. In addition, we have found that the non-extensivity of the system disappears as we approach $T\rightarrow0$. In future, we aim to study fluctuations and correlations of conserved charges on the nuclear and hyperonic matter in the realm of non-extensivity.

	\section*{Appendix}

The coupled equations  are obtained by minimizing the thermodynamic potential density, $\Omega_q$ with respect to the various fields of the $q$-PCQMF model and are given as

\begin{eqnarray}\label{sigma1}
	&&\frac{\partial \Omega_q}{\partial \sigma}= k_{0}\chi^{2}\sigma-4k_{1}\left( \sigma^{2}+\zeta^{2}
	+\delta^{2}\right)\sigma-2k_{2}\left( \sigma^{3}+3\sigma\delta^{2}\right)
	-2k_{3}\chi\sigma\zeta \nonumber\\
	&-&\frac{d}{3} \chi^{4} \bigg (\frac{2\sigma}{\sigma^{2}-\delta^{2}}\bigg )
	+\left( \frac{\chi}{\chi_{0}}\right) ^{2}m_{\pi}^{2}f_{\pi}- 
	\left(\frac{\chi}{\chi_0}\right)^2m_\omega\omega^2
	\frac{\partial m_\omega}{\partial\sigma}\nonumber\\
	&-&\left(\frac{\chi}{\chi_0}\right)^2m_\rho\rho^2 
	\frac{\partial m_\rho}{\partial\sigma}
	-\sum_{i=u,d} g_{\sigma}^i\rho_{q,i}^{s} = 0 ,
\end{eqnarray}
\begin{eqnarray}
	&&\frac{\partial \Omega_q}{\partial \zeta}= k_{0}\chi^{2}\zeta-4k_{1}\left( \sigma^{2}+\zeta^{2}+\delta^{2}\right)
	\zeta-4k_{2}\zeta^{3}-k_{3}\chi\left( \sigma^{2}-\delta^{2}\right)-\frac{d}{3}\frac{\chi^{4}}{{\zeta}}\nonumber\\
	&+&\left(\frac{\chi}{\chi_{0}} \right)
	^{2}\left[ \sqrt{2}m_{K}^{2}f_{K}-\frac{1}{\sqrt{2}} m_{\pi}^{2}f_{\pi}\right]-\left(\frac{\chi}{\chi_0}\right)^2m_\phi\phi^2 
	\frac{\partial m_\phi}{\partial\zeta}
	-\sum_{i=s} g_{\zeta}^i\rho_{q,i}^{s} \nonumber\\&=& 0 ,
	\label{zeta}
\end{eqnarray}
\begin{eqnarray}
	&&\frac{\partial \Omega_q}{\partial \delta}=k_{0}\chi^{2}\delta-4k_{1}\left( \sigma^{2}+\zeta^{2}+\delta^{2}\right)
	\delta-2k_{2}\left( \delta^{3}+3\sigma^{2}\delta\right) +\mathrm{2k_{3}\chi\delta
		\zeta} \nonumber\\
	& + &  \frac{2}{3} d \chi^4 \left( \frac{\delta}{\sigma^{2}-\delta^{2}}\right)
	-\sum_{i=u,d} g_{\delta}^i\rho_{q,i}^{s} = 0 ,
	\label{delta}
\end{eqnarray}
\begin{eqnarray}
	&&\frac{\partial \Omega_q}{\partial \chi}=\mathrm{k_{0}\chi} \left( \sigma^{2}+\zeta^{2}+\delta^{2}\right)-k_{3}
	\left( \sigma^{2}-\delta^{2}\right)\zeta + \chi^{3}\left[1
	+{\rm {ln}}\left( \frac{\chi^{4}}{\chi_{0}^{4}}\right)  \right]
	+(4k_{4}-d)\chi^{3}
	\nonumber\\
	&-&\frac{4}{3} d \chi^{3} {\rm {ln}} \Bigg ( \bigg (\frac{\left( \sigma^{2}
		-\delta^{2}\right) \zeta}{\sigma_{0}^{2}\zeta_{0}} \bigg )
	\bigg (\frac{\chi}{\mathrm{\chi_0}}\bigg)^3 \Bigg )+
	\frac{2\chi}{\chi_{0}^{2}}\left[ m_{\pi}^{2}
	f_{\pi}\sigma +\left(\sqrt{2}m_{K}^{2}f_{K}-\frac{1}{\sqrt{2}}
	m_{\pi}^{2}f_{\pi} \right) \zeta\right] \nonumber\\
	&-& \frac{\chi}{{\chi^2_0}}({m_{\omega}}^2 \omega^2+{m_{\rho}}^2\rho^2)  = 0,
	\label{chi}
\end{eqnarray}
\begin{eqnarray}
	\frac{\partial \Omega_q}{\partial \omega}=\frac{\chi^2}{\chi_0^2}m_\omega^2\omega+4g_4\omega^3+12g_4\omega\rho^2
	&-&\sum_{i=u,d}g_\omega^i\rho_{q,i}=0,
	\label{omega} 
\end{eqnarray}
\begin{eqnarray}
	\frac{\partial \Omega_q}{\partial \rho}=\frac{\chi^2}{\chi_0^2}m_\rho^2\rho+4g_4\rho^3+12g_4\omega^2\rho&-&
	\sum_{i=u,d}g_\rho^i\rho_{q,i}=0, 
	\label{rho} 
\end{eqnarray}
\begin{eqnarray}
	\frac{\partial \Omega_q}{\partial \phi}=\frac{\chi^2}{\chi_0^2}m_\phi^2\phi+8g_4\phi^3&-&
	\sum_{i=s}g_\phi^i\rho_{q,i}=0,
	\label{phi}  
\end{eqnarray}
\begin{eqnarray}
	\hspace*{0.4cm} 
	\frac{\partial \Omega_q}{\partial \Phi} =\bigg[\frac{-a(T)\bar{\Phi}}{2}-\frac{6b(T)
		(\bar{\Phi}-2{\Phi}^2+{\bar{\Phi}}^2\Phi)
	}{1-6\bar{\Phi}\Phi+4(\bar{\Phi}^3+\Phi^3)-3(\bar{\Phi}\Phi)^2}\bigg]T^4
	-\sum_{i=u,d,s}\frac{2k_BTN_C}{(2\pi)^3}
	\nonumber\\
	\int_0^\infty d^3k 
	\bigg[\frac{e_q\left(\frac{-(E_i^*(k)-{\nu_i}^{*})}{k_BT}\right)}{\left(1+e_q\left(\frac{-3(E_i^*(k)-{\nu_i}^{*})}{k_BT}\right)+3\Phi e_q\left(\frac{-(E_i^*(k)-{\nu_i}^{*})}{k_BT}\right)
		+3\bar{\Phi}e_q\left(\frac{-2(E_i^*(k)-{\nu_i}^{*})}{k_BT}\right)\right)^q}
	\nonumber\\
	+\frac{e_q\left(\frac{-2(E_i^*(k)+{\nu_i}^{*})}{k_BT}\right)}{\left(1+e_q\left(\frac{-3(E_i^*(k)+{\nu_i}^{*})}{k_BT}\right)
		+3\bar{\Phi} e_q\left(\frac{-(E_i^*(k)+{\nu_i}^{*})}{k_BT}\right)+3\Phi e_q\left(\frac{-2(E_i^*(k)+{\nu_i}^{*})}{k_BT}\right)\right)^q}\bigg] \nonumber \\=0,
	\label{Polyakov} 
\end{eqnarray}
and
\begin{eqnarray}
	\frac{\partial \Omega_q}{\partial \bar{\Phi}} =\bigg[\frac{-a(T)\Phi}{2}-\frac{6b(T)
		(\Phi-2{\bar{\Phi}}^2+{\Phi}^2\bar{\Phi})
	}{\mathrm{1-6\bar{\Phi}\Phi+4(\bar{\Phi}^3+\Phi^3)-3(\bar{\Phi}\Phi)^2}}\bigg]T^4
	-\sum_{i=u,d,s}\frac{2k_BTN_C}{(2\pi)^3}
	\nonumber\\
	\int_0^\infty d^3k 
	\bigg[\frac{e_q\left(\frac{-2(E_i^*(k)-{\nu_i}^{*})}{k_BT}\right)}{\left(1+e_q\left(\frac{-3(E_i^*(k)-{\nu_i}^{*})}{k_BT}\right)+3\Phi e_q\left(\frac{-(E_i^*(k)-{\nu_i}^{*})}{k_BT}\right)
		+3\bar{\Phi}e_q\left(\frac{-2(E_i^*(k)-{\nu_i}^{*})}{k_BT}\right)\right)^q}
	\nonumber\\
	+\frac{e_q\left(\frac{-(E_i^*(k)+{\nu_i}^{*})}{k_BT}\right)}{\left(1+e_q\left(\frac{-3(E_i^*(k)+{\nu_i}^{*})}{k_BT}\right)
		+3\bar{\Phi} e_q\left(\frac{-(E_i^*(k)+{\nu_i}^{*})}{k_BT}\right)+3\Phi e_q\left(\frac{-2(E_i^*(k)+{\nu_i}^{*})}{k_BT}\right)\right)^q}\bigg] \nonumber \\=0. 
	\label{Polyakov conjugate} 
\end{eqnarray}



\begin{thebibliography}{99}
	\bibitem{big} D. Boyanovsky et al., Ann. Rev. Nucl. Part. Sci. \textbf{56}, 441 (2006).
	
	\bibitem{dex2014} M. Buballa et al., J. Phys. G \textbf{41}, 123001
	(2014).
	
	\bibitem{baym}  G. Baym et al., Rep. Prog. Phys. \textbf{81}, 056902 (2018).
	
	\bibitem{li2019} B.-L. Li et al., Phys. Rev. D \textbf{99}, 043001 (2019).
	
	\bibitem{hinderer} T. Hinderer et al., Phys. Rev. D \textbf{81}, 123016 (2010).
	
	\bibitem{weber} F. Weber, Prog. Part. Nucl. Phys. \textbf{54}, 193 (2005).
	
	\bibitem{marty} R. Marty and J. Aichelin, Phys. Rev. C \textbf{87}, 034912 (2013).
	
	\bibitem{tsushima} K. Tsushima et al., Phys. Rev. C \textbf{59}, 2824 (1999).
	
	\bibitem{saito} K. Saito and A.W. Thomas, Phys. Lett. B \textbf{327}, 9 (1994).
	
	\bibitem{schaefer} B.J. Schaefer et al., Phys. Rev. D \textbf{76}, 074023 (2007).
	
	\bibitem{herbst} T.K. Herbst et al., Phys. Lett. B \textbf{696}, 58 (2011).
	
	\bibitem{menezes} D.P. Menezes et al., J. Phys. G \textbf{32}, 1081 (2006).
	
	\bibitem{stiele} R. Stiele et al., Phys. Lett. B \textbf{729}, 72 (2014).
	
	\bibitem{fukushima} K. Fukushima, Phys. Lett. B \textbf{591}, 277 (2004).
	
	\bibitem{ratii} C. Ratti, M.A. Thaler, W. Weise, Phys. Rev. D \textbf{73}, 014019 (2006).
	
	\bibitem{abu} M. Abu-Shady and H. M. Mansour, Phys. Rev. C \textbf{85}, 5 (2012).
	
	\bibitem{gatto} R. Gatto and M. Ruggieri, Phys. Rev. D \textbf{78}, 034015 (2011).
	
	\bibitem{schaefer2009} B.J. Schaefer and M. Wagner, Prog. Part. Nucl. Phys. \textbf{62}, 381 (2009).
	
	\bibitem{kovacs} P. Kovacs, Zs. Szep, and Gy. Wolf, Phys. Rev. D \textbf{93} 11, 114014 (2016).
	
	\bibitem{kovacs2022} P. Kovacs, G. Kovacs, and F. Giacosa, Phys. Rev. D \textbf{106 }11, 116016 (2022).
	
	\bibitem{wang} P. Wang et al., Commun. Theor. Phys. \textbf{36}, 71 (2001).
	
	\bibitem{wang2001} P. Wang et al., Nucl. Phys. A \textbf{688}, 791 (2001).
	
	\bibitem{manisha} M. Kumari and A. Kumar, Eur. Phys. J. Plus \textbf{136}, 19 (2021).

    \bibitem{nisha} N. Chahal and A. Kumar, Chin. Phys. C \textbf{46}, 6 (2022).
	
	\bibitem{tsallis} C. Tsallis, J. Stat. Phys. \textbf{52}, 479 (1998).
	
	\bibitem{zhao2020} Ya-Peng Zhao, Phys. Rev. D \textbf{101}, 096006 (2020).
	
	\bibitem{rozy2016} J. Rozynek and G. Wilk, Eur. Phys. A \textbf{52}, 13 (2016).
	
	\bibitem{bhat} T. Bhattacharyya et al., Eur. Phys. J. A \textbf{52}, 30 (2016).
	
	\bibitem{asmaa} Asmaa G. Shalaby, A. Phys. Pol. B \textbf{47}, 5 (2016).
	
	\bibitem{wilk}  G. Wilk and Z. Wlodarczyk, Eur. Phys. J. A \textbf{48}, 161 (2012).
	
	\bibitem{li} B.-C. Li, Y.-Z. Wang, and F.-H. Liu, Phys. Lett. B \textbf{725}, 352
	(2013).
	
	\bibitem{marques} L. Marques, J. Cleymans, and A. Deppman, Phys. Rev. D
	\textbf{91}, 054025 (2015).
	
	\bibitem{de} B. De, Eur. Phys. J. A \textbf{50}, 138 (2014).
	
	\bibitem{ryb} M. Rybczynski and Z. Wlodarczyk, Eur. Phys. J. C \textbf{74}, 2785 (2014).
	
	\bibitem{combe} G. Combe et al., Phys. Rev. Lett. \textbf{115}, 238301 (2015).
	
	\bibitem{tirnakli} U. Tirnakli and E. P. Borges, Sci. Rep \textbf{6}, 23644 (2016).
	
	\bibitem{cirto} L. J. L. Cirto, A. Rodriguez, F. D. Nobre et al., EPL \textbf{123}, 30003 (2018).
	
	\bibitem{phenix} A. Adare et al. (PHENIX Collaboration), Phys. Rev. C \textbf{83}, 064903 (2011).
	
	\bibitem{star} B. I. Abelev et al. (STAR Collaboration), Phys. Rev. C \textbf{75}, 064901 (2007).
 
	\bibitem{atlas} G. Aad et al. (ATLAS Collaboration), New J. Phys. \textbf{13}, 053033 (2011).
	
	\bibitem{alice} K. E. A. Aamodt (ALICE Collaboration), Eur. Phys. J. C \textbf{71},
	1655 (2011).
	
	\bibitem{cms} V. E. A. Khachatryan (CMS Collaboration), J. High Energy	Phys. \textbf{02}, 041 (2010).
	
	\bibitem{bediaga}  I. Bediaga, E.M.F. Curado, and J.M. de Miranda, Physica A
	\textbf{286}, 156 (2000).
	
	\bibitem{lavagno2000} W.M. Alberico, A. Lavagno, and P. Quarati, Eur. Phys. J. C \textbf{12}, 499 (2000).
	
	\bibitem{lavagno2007} A. Lavagno, A.M. Scarfone, and P.N. Swamy, J. Phys. A \textbf{40}, 8635 (2007).
	
	\bibitem{beck} C. Beck, Eur. Phys. J. A \textbf{40}, 267 (2009).
	
	\bibitem{wilk2000} G. Wilk and Z. Wlodarczyk, Phys. Rev. Lett. \textbf{84}, 2770 (2000).
	
	\bibitem{lavagno2001} A. Lavagno, P. Quarati, Phys. Lett. B \textbf{498}, 47 (2001).
	
	\bibitem{biro2005} T.S. Biro and G. Purcsel, Phys. Rev. Lett. \textbf{95}, 162302 (2005).
	
	\bibitem{biro2009} T.S. Biro, G. Purcsel, and K. Urmossy, Eur. Phys. J. A \textbf{40}, 325 (2009).
	
	\bibitem{lavagno2002} A. Lavagno, Phys. Lett. A \textbf{301}, 13 (2002).
	
	\bibitem{alberico2008}  W.M. Alberico et al., Physica A \textbf{387}, 467 (2008).
	
	
	\bibitem{quarati} P. Quarati and A.M. Scarfone, Astrophys. J. \textbf{666}, 1303 (2007).
	
	\bibitem{carvalho}  J.C. Carvalho et al., Astrophys. J. \textbf{696}, L48 (2009).
	
	\bibitem{livadiotis} G. Livadiotis and D.J. McComas, Astrophys. J. \textbf{714}, 971
	(2010).
	
	\bibitem{leubner} M.P. Leubner, Astrophys. J. \textbf{632}, L1 (2005).
	
	\bibitem{tsallis_2} C.  Tsallis, Introduction to nonextensive statistical
	mechanics: approaching a complex world (Springer Science, 2009).
	
	\bibitem{periera2007} F. I. M. Pereira, R. Silva, and J. S. Alcaniz
	Phys. Rev. \textbf{C} 76, 015201 (2007).
	
	\bibitem{rozy2009} J. Roynek and G. Wilk 2009 J. Phys. G: Nucl. Part. Phys. \textbf{36} 125108 (2009).
	
	\bibitem{cardoso} P.H.G. Cardoso, T. Nunes da Silva, and A. Deppman, et al., Eur. Phys. J. A \textbf{53}, 191 (2017).
	
	\bibitem{shen} Shen et al., Adv. High Energy Phys. 2017 \textbf{(2017)}, 4135329 (2017). 
	
	\bibitem{zhao2021} Ya-Peng Zhao et al., Chinese Phys. C \textbf{45}, 073105 (2021).
	
	\bibitem{zhao2023} Ya-Peng Zhao et al. Chinese Phys. C \textbf{47} 053103 (2023).
	
	\bibitem{alberico2000} W.M. Alberico, A. Lavagno, and P. Quarati, Eur. Phys. J. C \textbf{12}, 499 (2000).
	
	\bibitem{drago} A. Drago, A. Lavagno, and P. Quarati, Physica \textbf{A} 344 472477 (2004).
	
	\bibitem{lavagno2010} A. Lavagno et al, J. Phys. G: Nucl. Part. Phys. \textbf{37}, 115102 (2010).
	
	\bibitem{lavagno2011} A. Lavagno and D. Pigato, Eur. Phys. J. A \textbf{47}, 52 (2011).
	
	\bibitem{megias} E. Megias, D. P. Menezes, and A. Deppman, Physica \textbf{A} 421, 15 (2015).
	
	\bibitem{tiwari}  S.K. Tiwari et al. Eur. Phys. J. \textbf{C} 78, 938 (2018).
	
	\bibitem{schechter} J. Schechter, Phys. Rev. D \textbf{21}, 3393 (1980).
	
	\bibitem{gomm} H. Gomm et al., Phys. Rev. D \textbf{33}, 801 (1986).
	
	\bibitem{heide} E.K. Heide, S. Rudaz, and P.J. Ellis, Nucl. Phys. A \textbf{571}, 713 (1994).
	
	\bibitem{carter} G. Carter, P.J. Ellis, and S. Rudaz, Nucl. Phys. A \textbf{603 }, 367-386 (1996).
	
	\bibitem{ko} P. Ko and S. Rudaz, Phys. Rev. D \textbf{50}, 6877-6894 (1994).
	
	\bibitem{papa1999} P. Papazoglou et al., Phys. Rev. C \textbf{59}, 411 (1999).
	
	\bibitem{wang2003}  P. Wang et al., Phys. Rev. C \textbf{67}, 015210 (2003).
	
	\bibitem{wang2002} P. Wang et al., Nucl. Phys. A \textbf{705}, 455 (2002).
	
	\bibitem{wang2004}  P. Wang et al., Nucl. Phys. A \textbf{744}, 273 (2004).
	
	\bibitem{costa} P. Costa et al., Symmetry \textbf{2}, 1338 (2010).
	
	\bibitem{roessner} S. Roessner et al., Phys. Rev. D \textbf{75}, 034007 (2007).
	
	\bibitem{roessner2008} S. Roessner et al., Nucl. Phys. A \textbf{814}, 118 (2008).
	
	\bibitem{fukugita} M. Fukugita, M. Okawa, and A. Ukava, Nucl. Phys. B \textbf{337}, 181 (1990).
	
	\bibitem{gross2004} D. H. E. Gross, Physica A \textbf{340}, 76 (2004).
	
	\bibitem{gross2006} D. H. E. Gross, Physica A \textbf{365}, 138 (2006).
	
	\bibitem{li2013} B. C. Li, Y. Z. Wang, and F. H. Liu, Phys. Lett. B \textbf{725}, 352 (2013).
	
	\bibitem{cleymans2013} J. Cleymans et al., Phys. Lett. B \textbf{723}, 351 (2013).
	
	\bibitem{azmi} M. D. Azmi and J. Cleymans, J. Phys. G \textbf{41}, 065001 (2014).
	
	\bibitem{kodama} T. Kodama et al., 2005 Europhys. Lett. \textbf{70}, 439 (2005).
	
	\bibitem{garcia}  V. Garcia-Morales and J. Pellicer, Physica A \textbf{ 361}, 161 (2006).
	
	\bibitem{tewel} A. M. Teweldeberhan, A. R. Plastino, H. C. Miller, Phys. Lett. A \textbf{343}, 71 (2005).

    \bibitem{biro20052} T. S. Biro and A. Jakovac, Phys. Rev. Lett. \textbf{94}, 132302 (2005).
	
	\bibitem{khuntia} A. Khuntia et al., Eur. Phys. J. A \textbf{52}, 292 (2016).
%
	\bibitem{mao} H. Mao et al., J. Phys. G: Nucl. Part. Phys. \textbf{37}, 035001 (2010).
	
%
%
%
%
%
%
%
%
%
%
%
%
%
%
%
%
%
	
	
\end{thebibliography}
\end{document}